\documentclass[10pt, conference]{IEEEtran}
\IEEEoverridecommandlockouts
\usepackage{cite}
\usepackage{amsmath,amssymb,amsfonts}
\usepackage{algorithmic}
\usepackage{graphicx}
\usepackage{textcomp}
\usepackage{multirow}

\usepackage{adjustbox}

\usepackage{xcolor}
\usepackage{listings}

\usepackage{lipsum}

\usepackage[all=normal,leading,leadingfraction,paragraphs]{savetrees}

\usepackage[hyphens]{url}
\usepackage{hyperref}
\hypersetup{breaklinks=true}
\usepackage{soul}
\usepackage{color}
\usepackage{bm}
\usepackage{multirow}
\usepackage[anythingbreaks]{breakurl}

\definecolor{lightgray}{gray}{0.9}
\definecolor{LightCyan}{rgb}{0.88,1,1}

\newcommand{\lmttfont}{\fontfamily{lmtt}\selectfont}

\usepackage{ragged2e}
\usepackage{array}

\newcolumntype{L}[1]{>{\raggedright\arraybackslash}m{#1}}
\newcolumntype{C}[1]{>{\centering\arraybackslash}m{#1}}
\newcolumntype{R}[1]{>{\raggedleft\arraybackslash}m{#1}}

\lstdefinestyle{customc}{
  belowcaptionskip=1\baselineskip,
  breaklines=true,
  frame=L,
  xleftmargin=\parindent,
  language=C,
  showstringspaces=false,
  basicstyle=\scriptsize\ttfamily,
  keywordstyle=\bfseries\color{green!40!black},
  commentstyle=\itshape\color{purple!40!black},
  identifierstyle=\color{blue},
  stringstyle=\color{orange},
  numbers=left,
    stepnumber=1,
}

\def\BibTeX{{\rm B\kern-.05em{\sc i\kern-.025em b}\kern-.08em
    T\kern-.1667em\lower.7ex\hbox{E}\kern-.125emX}}
\begin{document}

\title{On Temporal Isolation Assessment in Virtualized Railway Signaling as a Service Systems}

\author{
    \IEEEauthorblockN{Domenico Cotroneo, Luigi De Simone, Roberto Natella}
    \IEEEauthorblockA{Università degli Studi di Napoli Federico II, Naples, Italy
    \\\{cotroneo, luigi.desimone, roberto.natella\}@unina.it}
}

\maketitle

\begin{abstract}

Railway signaling systems provide numerous critical functions at different safety level, to correctly implement the entire transport ecosystem. Today, we are witnessing the increasing use of the cloud and virtualization technologies in such mixed-criticality systems, with the main goal of reducing costs, improving reliability, while providing orchestration capabilities. Unfortunately, virtualization includes several issues for assessing temporal isolation, which is critical for safety-related standards like EN50128.
In this short paper, we envision leveraging the real-time flavor of a general-purpose hypervisor, like Xen, to build the Railway Signaling as a Service (RSaaS) systems of the future. We provide a preliminary background, highlighting the need for a systematic evaluation of the temporal isolation to demonstrate the feasibility of using general-purpose hypervisors in the safety-critical context for certification purposes.

\end{abstract}


\Urlmuskip=0mu plus 1mu

\section{Introduction}
\label{sec:introduction}


Railway signaling includes complex systems that aim to guarantee the safety of rail infrastructure, the environment, and human beings. It connects several sensors and other subsystems via signals, which allow synchronization and transmission of control data to different vital and non-vital components within the rail infrastructure \cite{5338991}. Generally, this infrastructure includes several modules on-board trains, whose task is to manage the safety of the train itself, keep the appropriate speed of the vehicle, and prevent collisions. 
Usually, rail interlocking plants are kept strictly close to tracks to be controlled and rely on complex architectures and several costly cabling and communication systems. These architectures forced railway providers to rent or buy buildings in specific locations to house them. 

Today, we are witnessing the use of cloud and virtualization technologies to develop a way of housing everything remotely, by implementing a \textit{Railway Signaling as a Service (RSaaS)} system. This new model will reduce dramatically operational and capital costs that companies must support; moreover, it results in innovations since rail signals can be easily controlled by exploiting smart sensors, which enable new ways of fault diagnosis, failure predictions, and predictive maintenance, increasing the overall transportation effectiveness \cite{rsaas_siemens, 9469907}. Besides railway, virtualization is increasingly used also in other industrial domains, such as avionic, automotive, Industrial Internet of Things (IIoT), and telco systems with the recent development of 5G \cite{shit2rail,windriveriot,hercules2020,5gcity_H2020,hermann2016design,klingensmith2019using, klingensmith2018hermes}. Commonly, industrial practitioners develop \textit{mixed-criticality systems}, which integrate functionalities at different safety and/or time-critical levels into common platforms to reduce the size, weight, power, and cost of hardware. This leads to numerous challenges, especially when adopting virtualization technologies, since safety, fault-tolerance, and real-timeliness must be guaranteed at the same time at different levels. Indeed, mixed-criticality systems have to satisfy stringent requirements imposed by safety-critical standards (e.g., \cite{do178b}, \cite{iso26262}, \cite{cenelec201150128}), which mostly refer to \textit{temporal} and \textit{spatial isolation} among software component. Spatial isolation includes the capability of isolating code and data between domains (e.g., virtual machines (VMs)), preventing tasks to alter private data belonging to other tasks. Instead, temporal isolation is about limiting the impact of resource consumption  (e.g., tasks running on a VM) on the performance of other software components (e.g., tasks running on the other VMs). Compared to spatial isolation, which is often guaranteed by only using hardware-assisted mechanisms (e.g., MMU), temporal isolation remains a critical and open research problem \cite{heiser2019can, ge2019time, cinque2022certify,  barletta2022achieving, cinque2021virtualizing}. To exacerbate the problem, the above-mentioned standards also recommend providing documentation about evidence of a \textit{fail-safe} and/or \textit{fail-stop} behavior for such systems, which ultimately prevent failures leading to human and cost losses.

In this short paper, we depict our ongoing work of developing an RSaaS. We describe a Proof of Concept (PoC) according to the classical fault-tolerant 2oo2 schema, along with a methodology for evaluating temporal isolation. The PoC is currently implemented on top of a real-time flavor of the Xen hypervisor and {\lmttfont PREEMPT\_RT}-patched guests, which are becoming increasingly a choice as a basis for safety-critical virtualized systems since it provides several features for supporting real-time applications along cloud computing support \cite{xen_fusa_sig}. Further, we highlight the need for a systematic assessment of temporal isolation by leveraging well-known Design of Experiments (DoE) \cite{park2007design} to rigorously estimate the impact and significance of the numerous existing factors in this kind of environment, against various operational scenarios. The objective is to provide guidelines to properly fine-tune these systems to provide temporal isolation evidence for certification purposes.

The rest of the paper is organized as follows. Section~\ref{sec:proposal} introduces the proposal. Section~\ref{sec:experimental_analysis} shows preliminary results. Section~\ref{sec:related} analyzes representative real-time hypervisors in the state of the art. Section~\ref{sec:conclusion} concludes the paper.

\section{Proposal}
\label{sec:proposal}

In this section, we provide the background needed for understanding the proposed RSaaS PoC. We highlight the peculiarities of the host side, guest side. We briefly introduce the need for temporal isolation assessment for certification purposes.

\subsection{Railway Signaling as a Service (RSaaS) PoC}

\begin{figure}[htbp]
    \centering
    \includegraphics[width=\columnwidth]{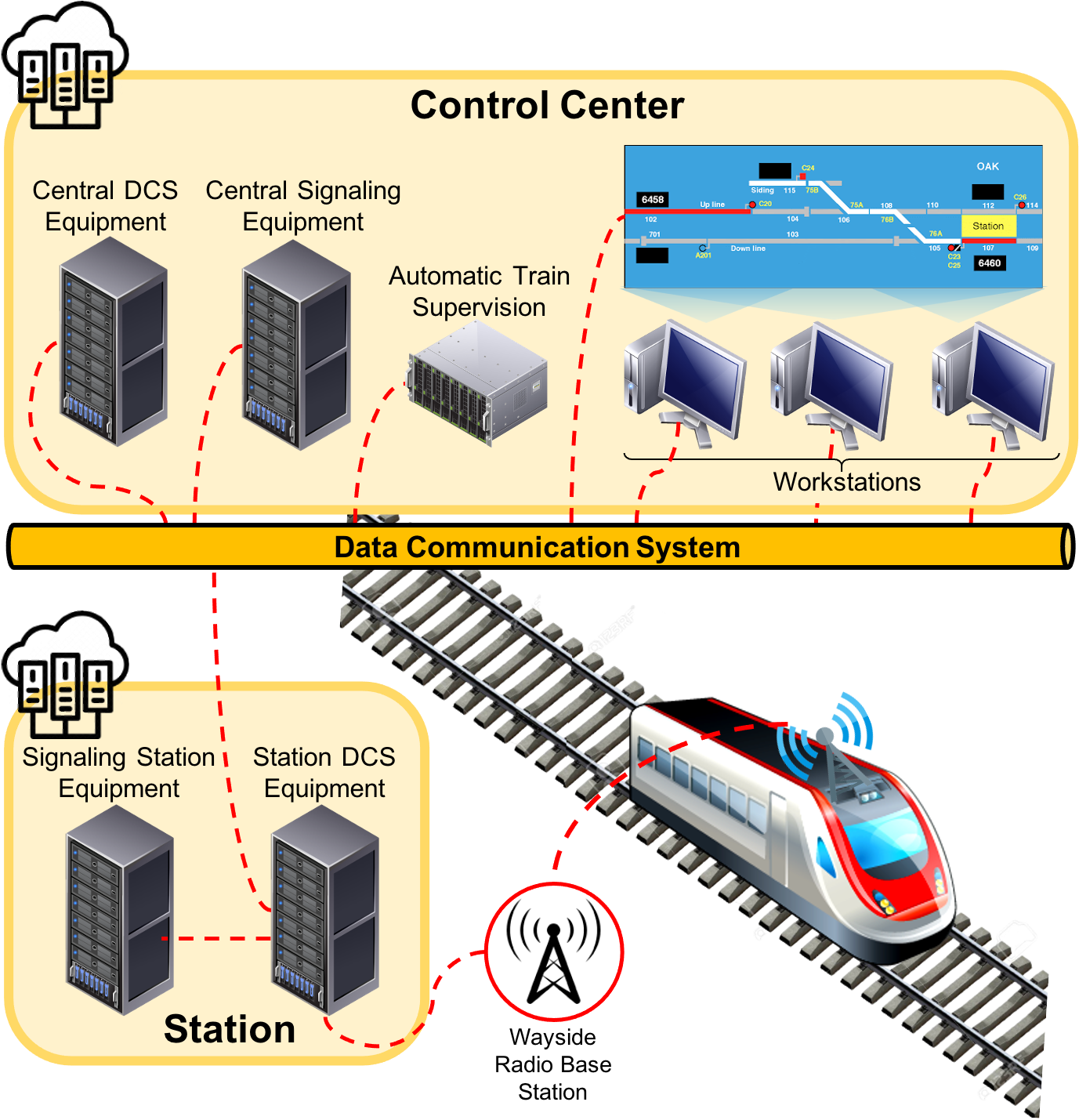}
    \caption{High-level architecture of a Communications Based Train Control (CBTC).} 
    \label{fig:signalin_as_a_service}
\end{figure}

In this work, we study the feasibility of developing a system that provides RSaaS by leveraging virtualization and cloud technologies. \figureautorefname{}~\ref{fig:signalin_as_a_service} envisions the high-level architecture, in which both the rail control center and station equipment are deployed in cloud data centers, replacing physical boards with several virtualized domains (e.g., virtual machines) running on top of high-performance real-time hypervisors. The data communication system (DCS) provides the communication between trains and ground carrying Communications Based Train Control (CBTC) signaling information and other telecommunication services onboard, like CCTV. The DCS network guarantees bidirectional, reliable, and secured exchange of data between equipment of the train control subsystems (e.g. zone controllers, interlocking, and carborne controllers). Wayside equipment is omitted for the sake of simplicity.

In general, signaling includes fail-safe applications, which require fail-stop approaches, with temporal monitoring of critical voting or time monitoring tasks. Indeed, CBTC signaling systems include several replicas for each component, which send the result of their measurements to a component, namely, the voter, which compares the results and eventually takes a decision by notifying if an error occurred or by acknowledging that the service is properly working. The most common configuration to guarantee fault-tolerance is the 2oo2 schema, in which there are 2 replicas for each system, the voter compares received values, and if they are different, an error is notified. When an error is notified, the train can be stopped and brought back to a fail-safe state. In order to properly handle common-mode errors, the 2oo2 Hot Redundancy (2oo2HR) schema is introduced. This schema considers two independent and isolated sections, i.e., a master section that is running by default, and a stand-by section that will be activated if an error occurs in the master section. 

\begin{figure}[htbp]
    \centering
    \includegraphics[width=.8\columnwidth]{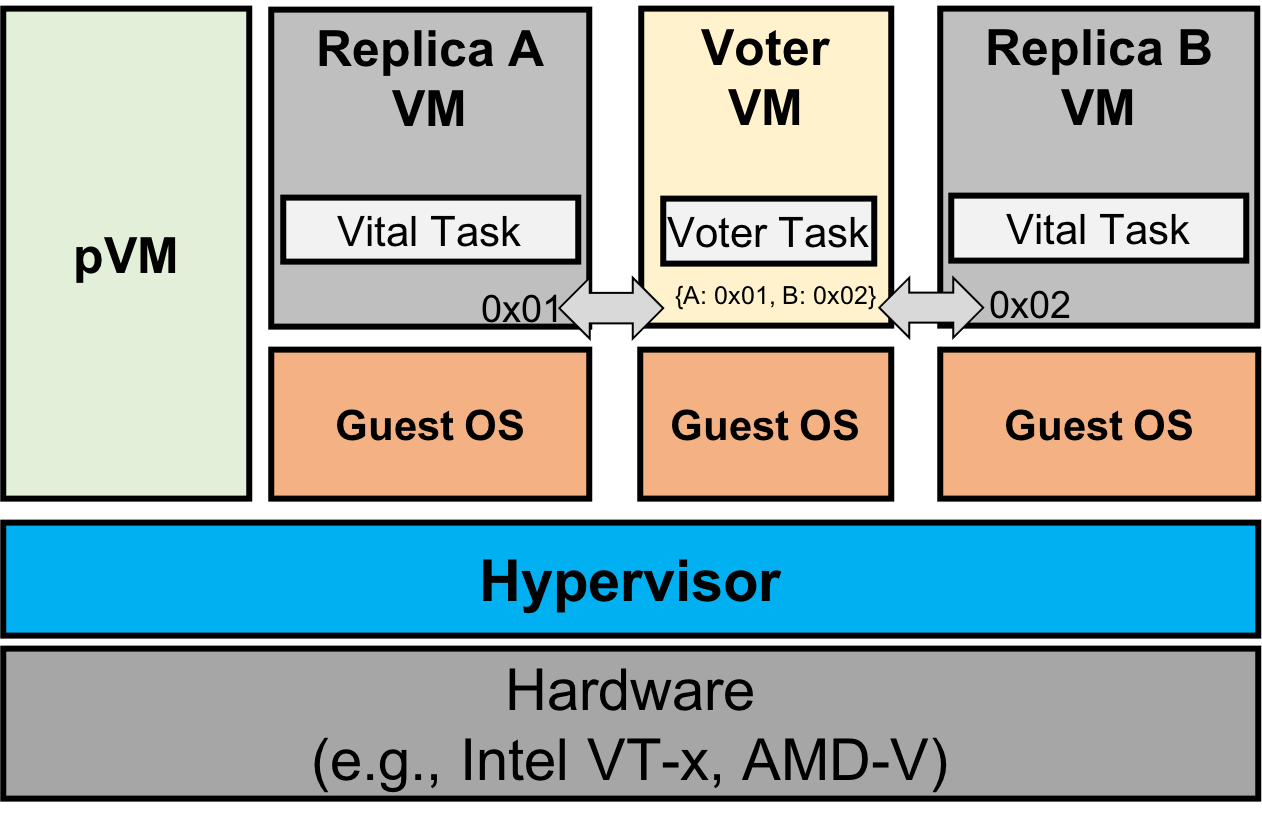}
    \caption{High-level PoC 2oo2 architecture.}
    \label{fig:PoC}
\end{figure}

According to the classical fault tolerance mechanism used in the railway domain, \figureautorefname{}~\ref{fig:PoC} shows the currently implemented PoC, which emulates a 2oo2 system, with 3 unprivileged virtual machines, two of them simulating the two replicas of a sensor, while one acting as a voter. The two replicas generate several values (simulating a measurement) over time, and they will send them to the voter. There is a probability that the two measurements will be different. When the voter receives a measurement from one of the two replicas will send an ACK to the sender. When both measurements will be received, the voter will compare them, returning a string that will indicate if the message received contain the same value.

The VMs are managed by a \textit{privileged VM} (pVM), running side by side with unprivileged VMs on top of a real-time hypervisor. In this short paper, we leverage the real-time flavor of the well-known Xen hypervisor for the pVM, and {\lmttfont PREEMPT\_RT}-patched Linux as a guest OS for unprivileged VMs.
\subsubsection{RT-Xen vs RTDS Xen}
\label{subsec:xen_bg}
The Xen hypervisor is an open-source type-1 hypervisor \cite{barham2003xen}, widely used as the basis for several commercial and open source applications, mainly involving server virtualization and recently applied in embedded appliances. In 2011, \textit{Xi et al.} \cite{6064510} proposed \textit{RT-Xen}. Based on vanilla Xen, RT-Xen provided schedulers for supporting soft real-time applications. Currently, only the \textit{Real-Time Deferrable Server (RTDS)} has been kept in the last version of RT-Xen (v2.2). Anyway, RTDS was included also in vanilla Xen starting from v4.5. RT-Xen v2.2 supports both \textit{Earliest Deadline First (EDF)} and \textit{Rate Monotonic (RM)} scheduling policies, and it is possible to switch on the fly between them; further, RT-Xen allows setting and getting scheduling parameters per-virtual CPUs (vCPUs). The RTDS scheduler assigns three parameters to every vCPUs, with a {\lmttfont budget} (the time allocated to a vCPU), a {\lmttfont period} (the period after which the budget can be refilled), and {\lmttfont extratime} (whether a vCPU can still run after budget expiration, when free physical CPU (pCPU) is available). When a vCPU starts running on a pCPU, its execution is not necessarily continuous. This means that its budget can be spent anytime during the period. A vCPU can execute if there is a free pCPU, or if it is running a vCPU with a lower priority. The deadline corresponds with the end of the period. RTDS can group pCPUs in \textit{CPU pools} and assigns each domain (i.e., VM) into one of these pools. For each CPU pool, RTDS keeps two queues: a \textit{global runqueue} which holds all the CPUs with a non-0 residual budget sorted by priority, and a \textit{depleted queue}, which holds all the CPUs that used up their budget.


\subsubsection{PREEMPT\_RT Linux}
\label{subsec:guest_side}

In the proposed PoC, we choose Linux as guest OS since it brings several advantages, like accessible and portable source code, in-depth research in the industrial field, creating a strong alternative also in embedded environments \cite{leppinen2017current}. The Linux Foundation supported several collaborative projects in the context of the real-time domain, developing the well-known {\lmttfont PREEMPT\_RT} patch \cite{reghenzani2019real}. This patch adds two more levels of preemption to the vanilla kernel, the \textit{Preemptible Kernel (Basic RT)} and the \textit{Fully Preemptible Kernel} options, in addition to the three levels provided by default. Further, {\lmttfont PREEMPT\_RT} removes most of the non-preemptible spinlocks,   replaces sleeping mutexes with {\lmttfont rt\_mutex} type that implements priority inheritance, and implements a new version of the \textit{high-resolution timers}. The {\lmttfont SCHED\_FIFO}, {\lmttfont SCHED\_RR} (round-robin), and the {\lmttfont SCHED\_DEADLINE} (implementation of the \textit{Earliest Deadline First (EDF)}) are the real-time scheduler policies implemented in Linux. For both {\lmttfont SCHED\_RR} and {\lmttfont SCHED\_FIFO} the user needs to statically assign priorities. Instead, {\lmttfont SCHED\_DEADLINE} leverages \textit{runtime}, \textit{period}, and \textit{deadline} parameters to schedule tasks. {\lmttfont SCHED\_DEADLINE} allocates to a task \textit{runtime} $\mu$s of execution time every \textit{T} $\mu$s specified with the \textit{period}. This allocated time is available within the specified \textit{deadline} from the beginning of the period.

\subsection{Temporal Isolation Assessment}

In the context of virtualization, temporal isolation is the ability to isolate or limit the impact of resource consumption (e.g. CPU, network, disk) of a VM on the performance degradation of other VMs and even against the host. This means that a critical task running in a VM must not cause delays to other critical and non-critical tasks running in different VMs, avoiding phenomena such as starvation, reduced throughput, and increased latency. Indeed, we need to take into account several knobs, e.g., guest OSes schedulers, vCPUs/pCPUs mapping, hypervisor scheduler, and so on: if not properly managed they could easily break temporal isolation. 

In this work in progress paper, we highlight the need of performing systematic experiments to unveil potential interference under corner case conditions (e.g., stressful privileged/unprivileged VMs, faults in the hypervisor, bad configurations, etc.). The ultimate objective is twofold: i) quantifying how strong is the temporal isolation assured by the hypervisor, and ii) providing guidelines for practitioners to fine-tune the system parameters in case of weak isolation. According to EN50128, we identified, among others, critical requirements like \textit{D.45 Response Timing and Memory Constraints}, \textit{D.3 Avalanche/Stress Testing}, and \textit{D.25 SEEA – Software Error Effect Analysis} that should be considered to provide proper temporal isolation, even in lower SIL levels. The main idea is to leverage the well-known DoE to rigorously assess the impact and significance of system parameters in different operational scenarios. 

\section{Preliminary Results}
\label{sec:experimental_analysis}

At the current state of the work, we assess temporal isolation by implementing a \textit{stress VM}, with different configurations, that runs side-by-side with replicas VMs, voter VM, and pVM depicted in \figureautorefname{}~\ref{fig:PoC}. For the sake of simplicity, the voter is implemented naively, i.e., it receives measures from replicas and just compare without alerting replicas. Further, the purpose of the stress VM is to cause a high computational load that likely results in interference between the existing VMs, in order to comply with the \textit{D.3 Avalanche/Stress Testing} EN50128 requirement. We leveraged DoE to perform experiments to precisely analyzed what is the degree of induced interference (e.g., increasing variance of latencies) for both different stress VM vCPU/pCPU mappings and guest OSes real-time scheduler policies, by fixing the Xen RTDS parameters. 
The PoC is preliminarily deployed on a hardware platform with 8 pCPUs Intel(R) Xeon(R) E5-2667, 16GB RAM, 200GB HDD, using Xen v4.13 as the hypervisor. We set up Xen RTDS scheduler with the default parameters, i.e., $4000 \mu s$ for budget and $10000 \mu s$ for the period, and set the vCPU/pCPU mappings for the pVM (i.e., Xen Dom0), replicas VM, voter VM, and stress VM as described in the following:

\begin{itemize}

    \item \textit{pVM}: 2 vCPU pinned on {\lmttfont pCPU0} and {\lmttfont pCPU1};
    
    \item \textit{voter VM}: 2 vCPU pinned on physical {\lmttfont pCPU2} and {\lmttfont pCPU3};
    
    \item \textit{replicas VM}: 1 vCPUs pinned respectively on the {\lmttfont pCPU4} and {\lmttfont pCPU5};
    
    \item \textit{stress VM}: it can freely use the {\lmttfont pCPU6} and {\lmttfont pCPU7} as described in the next.
    
\end{itemize}

The idea behind the vCPU/pCPU configuration for the stress VM is to consider 3 load levels, i.e., \textit{Low}, \textit{Mid}, and \textit{High}. In particular:

\begin{enumerate}

    \item \textit{Low}, includes only 1 vCPU allocated to the stress VM, and it is pinned on one of the free pCPUs.
    
    \item \textit{Mid}, includes 4 vCPUs allocated to the stress VM. Specifically, 2 vCPUs are pinned on the 2 free pCPU, while the other 2 vCPUs are pinned on the pCPU0 and pCPU1. In this way, the stress VM will use the pCPUs already allocated to the pVM, and will not interfere with the execution of the replicas and voter VMs.
    
    \item \textit{High}, includes 4 vCPUs allocated to the stress VM, and all of them are pinned on the pCPUs used by the replicas and the voter. In this scenario, we are increasing the probability of interference between stress and replicas and voter VMs.
    
\end{enumerate}

The other parameter on which we can operate is the scheduler policy for the guest OSes used for replicas and the voter VMs. In these tests, we set VMs only with the {\lmttfont SCHED\_FIFO} policy. Hence, our test plan contemplates two factors, i.e., the \textit{Stress load} and the \textit{guest OS Scheduler}, with three and one levels respectively. This means that a full factorial Design of Experiment (DoE) establishes 3 experiments to be performed. 


For each configuration, we set $N = 35$ measurements that each replica send to the voter both w/ stress and w/o stress scenarios. For each experiment, \tablename~\ref{tab:results} shows the results in terms of the average and standard deviation of latencies obtained in the scenario without the stress workload, and the increase of the same metrics when the stress workload is activated. Further, in order to understand if the difference between the latency distributions obtained in both w/o and w/ scenarios is statistically significant, we use the \textit{t-test} by showing the obtained \textit{p-value}. We set a p-value threshold equal to $0.05$ to consider the two distributions statistically different or not.

\begin{table}
\centering
\caption{Results for tested configurations.}
\label{tab:results}
\begin{adjustbox}{max width=\columnwidth}
\begin{tabular}{c|c|cc|cc|c}
\hline \hline                
\multirow{2}{*}{\textbf{Cfg. ID}} & \multirow{2}{*}{\textbf{VM}} & \multicolumn{2}{c|}{\textbf{w/o stress}} & \multicolumn{2}{c|}{\textbf{w/ stress}}     & \multirow{2}{*}{\textbf{\begin{tabular}[c]{@{}c@{}}t-test\\ p-value\end{tabular}}} \\ 
                                  &                              & \multicolumn{1}{c|}{\textit{Avg.} [$\mu$s]}     & \textit{SD}       & \multicolumn{1}{c|}{\textit{Avg. incr.}} & \textit{SD incr.} &                                                                                    \\ \hline
\multirow{2}{*}{LOW\_FIFO}             & Replicas                     & \multicolumn{1}{c|}{410.33}   & 114.16   & \multicolumn{1}{c|}{22.47\%}    & 22.72\%   & 8.31E-10                                                                           \\
                                  & Voter                        & \multicolumn{1}{c|}{763.97}   & 44.28    & \multicolumn{1}{c|}{19.46\%}    & 154.31\%  & 8.31E-10                                                                           \\ \hline
\multirow{2}{*}{MID\_FIFO}             & Replicas                     & \multicolumn{1}{c|}{392.19}   & 60.88    & \multicolumn{1}{c|}{99.16\%}    & 1215.49\% & 8.50E-05                                                                           \\
                                  & Voter                        & \multicolumn{1}{c|}{862.24}   & 65.40    & \multicolumn{1}{c|}{78.76\%}    & 1373.85\% & 6.19E-05                                                                           \\ \hline
\multirow{2}{*}{HIGH\_FIFO}             & Replicas                     & \multicolumn{1}{c|}{371.26}   & 48.18    & \multicolumn{1}{c|}{106.40\%}   & 1762.52\% & 3.39E-04                                                                           \\
                                  & Voter                        & \multicolumn{1}{c|}{750.00}   & 70.87    & \multicolumn{1}{c|}{109.53\%}   & 1806.65\% & 7.39E-04                                  \\                        \hline \hline                
\end{tabular}
\end{adjustbox}   
\end{table}

As we expected, preliminary results show that the vCPU/pCPU mapping is a highly significant factor for temporal interference since exists a non-negligible increase in the average latencies and standard deviation in scenarios with stress. For interlocking systems, there is no standard agreement on performance requirements in Europe, and each railway infrastructure can adopt its own thresholds. For example, in \cite{RFI/DTC/DNS.SS.IM/009/041} the requirement for acceptable voting response time is between $350ms$ and $500ms$. Despite the obtained results appearing to be better according to the abovementioned thresholds, the voting logic adopted in this preliminary work is not representative of real operational conditions; rather, the most critical point to take into account is the high variability of these latency times. In future work, we planned another set of experiments to understand if different configurations of the hypervisor scheduler can mitigate temporal interference, even under stressful conditions.



\section{Related Work}
\label{sec:related}
\textit{PikeOS} \cite{pikeos} is a commercial hypervisor from SYSGO, used also in the railway domain. PikeOS supports multi-core platforms and its architecture is based on the L4 microkernel. PikeOS has been the target and the basis of several academic and industrial evaluations. Regarding certification and formal verification, in \cite{verbeek2015formal} the authors formalized the hardware-independent security-relevant part of PikeOS to prove intransitive noninterference properties \cite{roscoe1999intransitive}; moreover, in  \cite{baumann2009verifying} the authors presented first results in the verification of the PikeOS microkernel system calls. \textit{Jailhouse} \cite{jailhouse} is a Linux-based partitioning hypervisor developed by Siemens, enabling AMP applications to cooperate with the Linux kernel to run bare-metal applications or properly configured guest OSes. The main purpose is to enhance isolation rather than provide classic virtualization; thus, the hypervisor splits CPUs, memory, I/O ports, PCI devices, etc., into \textit{cells}, i.e., strongly isolated domains, each of them assigned to one guest OS, and its applications, called \textit{inmates}. Recently, in the context of the Hercules H2020 project, page coloring support has been added to Jailhouse, showing promising results about cache isolation. However, temporal isolation evidence is not documented. Similarly, \textit{Bao} \cite{martins2020bao} is a lightweight bare-metal partitioning hypervisor, designed ad-hoc for mixed-criticality IoT systems. Bao focuses on security and safety requirements by providing strong isolation, fault-containment, real-time features, and implements page coloring \cite{cache_coloring}. Bao has been recently ported to Xilinx’s ZCU104 MPSoC, showing reduced overheads while keeping low partition interference. \textit{Xtratum} \cite{masmano2009xtratum} is a para-virtualized partitioning hypervisor for the avionic domain, built according to the ARINC 653 standard. Xtratum supports several CPU families (e.g., Intel, Leon, ARM), providing strong temporal isolation between virtual partitions by exploiting a fixed cyclic scheduler. Spatial separation is guaranteed by imposing partitions allocated statically at different physical memory addresses, with no memory areas shared between partitions. Xtratum defines a minimum set of deterministic hypercalls and enables interrupts only for currently running partitions, to minimize temporal interference. Cache coherence is guaranteed by setting memory areas as \textit{non-cacheable} in the virtual memory map, unless target platforms implement hardware virtualization features or cache coherency acceleration.

In this short paper, we focused on Xen since we are assisting a proliferation of projects that leverage it as a basis for safety-critical virtualized systems. Xen currently provides real-time support for scheduling (ARINC, RTDS, Null schedulers), a minimal size (less than 30KSLOC) for ARM-based hardware environments, paravirtual and GPU mediation for rich I/O, TEE virtualization support, and \textit{Dom0less} architecture \cite{xen_fusa_sig}. Xen is also gaining popularity in the automotive domain, thanks to vertical initiatives like the Automotive Grade Linux (AG according to ISO 26262 {\cite{agl_kvm}}. Finally, Xen fully supports cloud computing infrastructures, providing mechanisms that include migration, balancing, and high availability. Thus, it can easily support potential solutions for orchestrating mixed-criticality scenarios like RSaaS, by running RTOSes and GPOSes consolidated on the same hardware platform. In that direction, Xilinx is currently developing a lightweight solution called \textit{RunX}, which exploits Xen to both run containers as VMs in the context of the industrial IoT and edge devices, which often require assuring real-time behavior \cite{lf_edge}. It is worth noting that the problem of time partitioning is still left exposed for multi/many cores architecture.

\section{Conclusion and Future plan}
\label{sec:conclusion}

In this short paper, we presented ongoing research on using a real-time flavor of general-purpose hypervisors for orchestrating virtualized signaling railway systems. In particular, we highlighted the need for a methodology intending to evaluate temporal isolation property for certification purposes. The proposal includes the use of DoE for planning experiments to \textit{i)} properly quantify isolation provided by the chosen hypervisor and \textit{ii)} providing guidelines to carefully tuning crucial parameters like guest OS and hypervisor schedulers, vCPU/pCPU mapping, etc. These, are some main levers on which acting to mitigate temporal interference. Future plans include applying the proposed approach against other configurations of Xen, like using partitioning (i.e., the \textit{null scheduler}) and Dom0-less features, and using different real-time communication approaches (e.g., XDP \cite{xdp}) for unprivileged domains.


\section*{Acknowledgments}
This work has been supported by the project COSMIC of UNINA DIETI.

\IEEEtriggeratref{14}

\bibliographystyle{IEEEtran}
\bibliography{bibliography}

\end{document}